\documentclass{xray_symp}
\usepackage{graphics,psfig}
\def\R{~ROSAT}

\def\n6240plots{/users/papade/STARBURST/NGC6240/PLOTS/}
\def\vv{\vspace*{-0.27mm}}
\def\etal{et\ al.}
\def\sbu{${\rm mag\,\,arcsec^{-2 }}$\ }
\def\xbu{${\rm counts\,\,ksec^{-1}\,\,arcmin^{-2}}$\ }

\def\ergsec{${\rm erg\,\,s^{-1}}$}
\newcommand{\mima}[2]{$_{\scriptscriptstyle -#1}^{\scriptscriptstyle +#2}$}

\def\hi{H\,{\small I}}
\def\ha{H$\alpha$}
\def\hn2{H$\alpha$+{[}N{\small II}{]}\  }

\def\lx{$\log(L_{\rm X})$}

\def\lxb{$\log(L_{\rm X}/L_{\rm B}$)}

\def\lirb{$\log(L_{\rm FIR}/L_{\rm B}$)}
\begin{document}
\title{X-rays from Interacting and Merging Galaxies}
\author{K.J.\ Fricke \and P.\ Papaderos}  
\institute{Universit\"{a}ts-Sternwarte G\"{o}ttingen, 
Geismarlandstra\ss e 11, 37083 G\"{o}ttingen, Germany}
\maketitle

\begin{abstract}
We have used \R\ PSPC and HRI to study the properties of the soft
X-ray emission in a sample of colliding starburst galaxies. 
Together with data from other spectral ranges (optical, radio and FIR) we 
investigate the emergence of soft X-ray emission during progressive
stages of interaction and merging. 
\end{abstract}
\section{Introduction}
Interactions and merging of galaxies must be considered major drivers of 
galaxy evolution (morphology, spectral evolution, gas-dynamics, high-energy 
processes, AGN formation). A merger sequence for the formation of elliptical
galaxies from interacting spirals was first proposed by Toomre\ (1977) based 
on numerical simulations. 
Inclusion of gas dynamics and star formation using Tree-SPH codes is now state 
of the art\,(cf. Mihos \& Hernquist 1996) and allows a variety of predictions 
for the induced star formation and dynamical phenomena to be compared with 
observations. 
In Sect.\,2 one of these systems, NGC\ 6240, is discussed in detail 
with respect to spatial structure, X-ray emission,  
optical colors, FIR-, and H$\alpha$ emission. 
Then we discuss the X-ray properties and X-ray evolution of a sample
representing interacting and merging systems along the merger sequence
and we look for systematic changes of the luminosities
$L_{\rm X}$, $L_{\rm B}$, $L_{\rm FIR}$, and the contribution 
of extended X-ray emission along that sequence.
\section{NGC\ 6240}
This system is in the stage of violent merging from the evidence of its
extreme FIR luminosity (Soifer et al. 1984) and peculiar morphology 
in the optical (Vorontsov--Vel'yaminov 1977) 
(cf. Fig.\ 1, this paper).  
Further evidence for its merger interpretation has been provided by the
detection of two centrally located nuclei separated in I and r images
by 1\farcs 8 (Fried \& Schulz 1983) and with a relative velocity of 
150~${\rm km\, s^{-1} }$ (Fried \& Ulrich 1985).
The double nucleus of this violent merger is also 
clearly seen on a filtered B-band image (Fig.\ 1). 
Evidence for shocked gas in the region between either nucleus
has been provided by the detection of strong H$_2$ 2.12$\mu$m emission
(van der Werf \etal\ 1993) which is consistent with 
the LINER nature diagnosed from the optical spectrum (Fosbury \& Wall 1979).

The merging in NGC\ 6240 is accompanied by a strong nuclear starburst
particularly showing up in the $2.3\mu {\rm m}$ vibrational absorption features 
arising from a substantial population of red supergiants 
(Rieke \etal\ 1985, Lester \etal 1988, van der Werf et al. 1993). 

The morphologically complex \hn2\ emission pattern of 
NGC~6240 with loops extending out to 60~kpc and
a total luminosity of $1.7\times 10^{42}$ \ergsec\ 
(Heckman \etal\ 1987) witnesses a starburst--driven 
galactic superwind.
On smaller scales ($\la 8$ kpc) effects of these strong gas motions, symmetrically
to the nuclei, are seen in maps of the H$\alpha$--equivalent width 
(Fig.\ 3,left). In this region 
enhanced emission from ionized gas is detected in the 
U--B map (Fig.\ 3,middle).

\R\ PSPC maps (Fricke \& Papaderos 1996) reveal a nuclear source 
accompanied by a very extended (2\arcmin) and one-sided soft source 
towards the West very closely following the H$\alpha$ maps of Heckman 
et al. (1987, see above).

The soft (0.1-2.4 keV) emission of NGC\ 6240 can be described by a thermal bremsstrahlung
model with a temperature kT=(0.8\mima{0.3}{0.5}) keV and an intrinsic absorbing column 
density of (1.1\mima{0.7}{1.2})$\times 10^{21}$ cm$^{-2}$. 
The same model yields an intrinsic luminosity of (3.1\mima{0.9}{1.2})$\times 10^{42}$ \ergsec.

Integration of radial intensity profiles of the PSPC-data yield
a $\sim 40$\% contribution of the extended emission 
to the 0.1--2.4 keV luminosity of the source. 
Furthermore, the extended nature of the X-ray emission in NGC\ 6240 is 
evident from the HRI-map superposed on the R--band image 
(Fig.\,1,middle). 
In the right panel of the same figure the intensity distribution of 
the extended HRI emission reaching out to $\sim 50$\arcsec\ is compared 
to the point response function (PRF) of the HRI and to the H$\alpha$ intensity distribution. 
\begin{figure*}[t]
\begin{picture}(16.8,5.5)
\put(12,0){\psfig{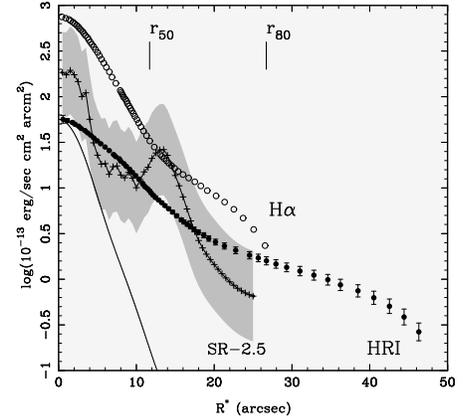}}
\end{picture}
\caption[]{{\bf left:} B-band exposure of NGC\ 6240. Contours, corrected for galactic
extinction (A$_{\rm B}$=0.29 mag) are shown at the intensities from
20.5 to 26\ \sbu in steps of 0.5 mag. 
Either nucleus is visible in the inset to the upper-right, displaying  
a filtered map of the central 12\arcsec$\times$12\arcsec\ region of the system. 
{\bf middle:} \R\ HRI exposure of NGC\ 6240 overlaid to a B-band exposure. 
The outermost contour corresponds to an intensity 7 \xbu\ above the background.
{\bf right:} Radial intensity profiles of the X-ray emission derived from \R\ HRI maps. 
The radii encircling 50\% and 80\% of the X-ray flux of the source were determined to
r$_{50}$=11\farcs 7 and r$_{\rm 80}$=26\farcs 7, i.e. they are 
considerably larger than the radii r$_{50}$=3\farcs 3 and r$_{80}$=5\farcs 9 
of the on-flight instrumental PRF (solid curve). 
This diagram shows that NGC\ 6240 is an extended X-ray source in agreement with the
result by Fricke \& Papaderos (1996).
The shaded region displays the radial distribution of the HRI softness ratio 
SR (cf. Wilson et al. 1992). 
The \ha-intensity distribution is shown by the open circles.}
\label{IG1}
\end{figure*}
%
The extended emission of $\sim 1.2\times 10^{42}$ \ergsec\ 
(40\% of the total soft X-ray emission) can 
naturally be explained by a starburst wind ISM/halo interaction.
Decomposition of surface brightness profiles (cf. Fig.\ 2) yields an absolute
B-magnitude of --21.2 mag for the starburst component of NGC\ 6240.
With this luminosity, from the models of Leitherer \& Heckman (1995), and adopting 
a burst duration of $\sim 25$ Myr we obtain a {\it mechanical} energy output 
from the starburst of $\sim 10^{43}$ \ergsec. 
The Suchkov et al. (1994) starburst wind model invoking a shocked ISM/halo 
then allows to account for the 40\% diffuse X-ray emission 
inferred above.
\begin{figure}[!br]
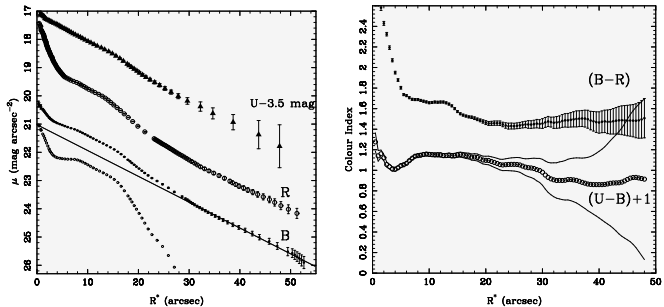

\begin{picture}(8,4)
\put(0,0){\psfig{figure=ngc6240f4.ps,height=4cm,angle=-90.,clip=}}
\put(4.5,0){\psfig{figure=ngc6240f5.ps,height=4cm,angle=-90.,clip=}}
\end{picture}
\caption[]{(a) Surface brightness profiles of NGC\ 6240 in U, B and R. Note the outer 
exponential intensity decrease for radii $>30$\arcsec. Profile decomposition into 
an exponential component (solid line) and the excess luminosity (small open circles)
is shown for the B-band profile.
(b) Radial B--R and U--B colour distribution. 
For radii $\la 6$\arcsec\ the B--R profile shows a gradient of 0.4 mag\ kpc$^{-1}$ and attains 
a central value $\ga 2.6$ mag. 
This reddening is not due to superimposed \ha-Emission which contribution was estimated to 
$\sim 6$\% to the R-band luminosity within the central region but instead of intrinsic 
absorption.}
\label{IG2}
\end{figure}
This starburst interpretation of the extended X-ray emission in NGC\ 6240 
is not to be confused with the situation at higher energies.
ASCA-observations (Iwasawa \& Comastri 1998) 
indicate that for an energy $>3$ keV 
the X-ray emission of NGC\ 6240 is dominated by reflection of photons
from a hidden AGN obscured by at least a $2\times 10^{24}$ cm$^{-2}$ 
column density of molecular gas.
This is consistent with the evidence from the B--R map (Fig.\ 3,right) 
and the B--R profile (Fig.\ 2) of
steadily increasing colours towards the nuclear region. 
%
\section{Correlations of Observables for a dynamical sequence}
%
Recently a first attempt has been made to discuss HRI and PSPC data for an 
{\it evolutionary sample} of 8 interacting/merging galaxies along essentially 
the Toomre\,(1977) merging sequence (Read \& Ponman 1998). 
We discuss here a more extended sample of 22 interacting systems 
representing widely different dynamical stages, ranging from weak 
gravitational interactions through close encounters to completely 
merged systems. 
Included in this sample are the two objects NGC\ 2623 and AM\ 1146-270 
from Read \& Ponman (1998) and the objects Arp\ 284, NGC\ 7252 
and NGC\ 6240, thoroughly discussed by us elsewhere (Papaderos \& Fricke 1998, 
Fritze-v. Alvensleben et al. 1998, this paper). 
Besides the X-ray luminosities $L_{\rm X}$, we determined from surface 
photometry of PSPC-maps the ratios of extended to pointlike emission, e/p.

A tight correspondence is found between \lxb\ and 
\lx\ (Fig.\,4a) and, less pronounced, for 
\lxb\ vs. \lirb\ (Fig.\,4b). 
\begin{figure*}[!t]
\begin{picture}(16.8,7)
%
\end{picture}
\caption[]{{\bf left:} \ha-equivalent width EW(\ha) map of the central region 
of NGC\ 6240 (D=98 Mpc). The overlayed contours, computed from a filtered B-band 
image, delineate the complex morphological structure of the merging system.
The EW(\ha) shows two enhancements symmetrically to the nuclei where
it attains values $\ga 260\,{\rm \AA}$. At the nuclear region 
EW(\ha) decreases to $\la 180\,{\rm \AA}$. 
The EW(\ha) morphology is most likely not due to the 
decreasing surface brightness of the stellar continuum in the vicinity of the
nuclei but witnesses a kinematically perturbed warm gas component.
The position of the nuclei is indicated in the inset (upper-right) by crosses.
{\bf middle:} U--B map of NGC\ 6240. The overlayed contours have the same meaning as 
in the diagram to the left. The reddest regions ($>0.3$ mag) are aligned with the 
major axis of the system between 4 and 8 kpc from the nuclear region.
The regions indicated by the circles display blue 
U--B colours ranging between --0.3 mag (region b) and --0.6 mag (region a).
Such colours, being much bluer than the average U--B index of $\sim +0.1$ mag 
for radii $>10$\arcsec\ (cf. Fig.\,2), point to the presence of strong emission by 
ionized gas, especially {[}OII{]}$\lambda\lambda 3726,3729$, in the vicinity of 
the nuclear region of NGC\ 6240,
in agreement with the EW(\ha)-morphology (cf. Barbieri et al. 1993).
{\bf right:} The diffuse B--R colour distribution does not show any correspondence 
to the optical morphology (contour lines) and attains values $\ga 2.6$ mag towards the nuclear 
region (cf. Fig.\,2). This indicates the increasing extinction within NGC\ 6240.}
\label{IG3}
\end{figure*}
Such relations support the view that the enhanced X-ray emission 
in merging systems is closely linked to the strength of the
induced starburst activity. 
The development of extended emission,
with the e/p-ratio being typically $>0.5$, seems to be 
a common feature of all mergers. 
No evidence for a relation between e/p and \lirb\ is 
observed (Fig.\,4c). 
\begin{figure}
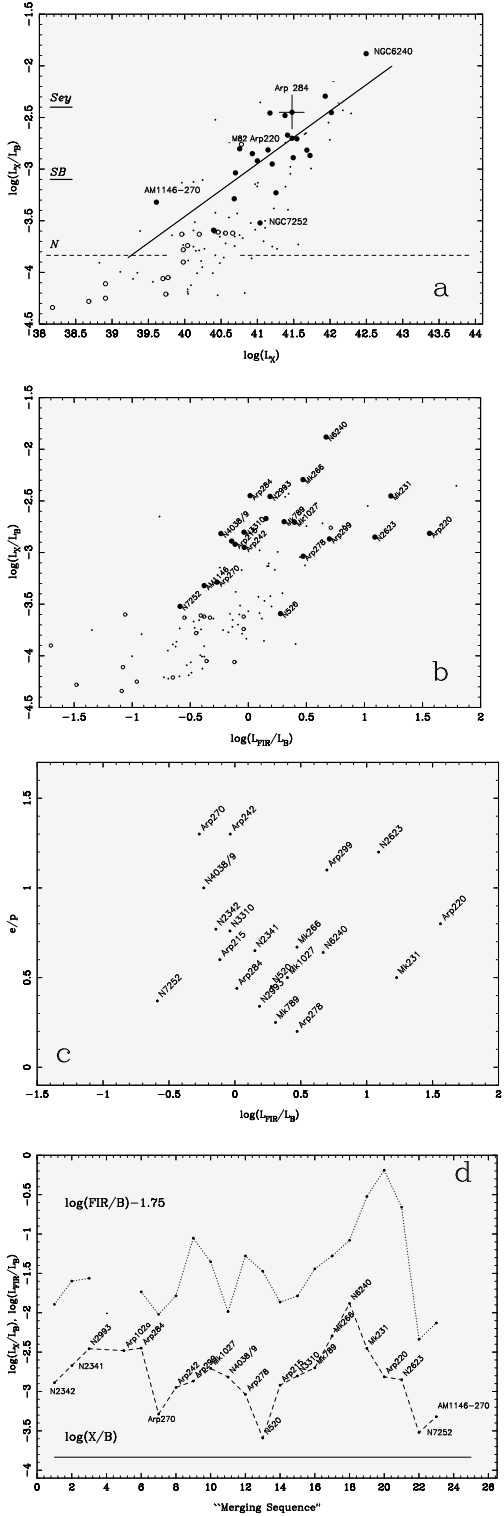

\begin{picture}(7,20)
\put(1,15.2){\psfig{figure=ig1.ps,height=4.8cm,angle=-90,clip=}}
\put(1,10.1){\psfig{figure=ig2.ps,height=4.8cm,angle=-90,clip=}}
\put(1,5.1){\psfig{figure=ig3.ps,height=4.8cm,angle=-90,clip=}}
\put(1,-0.1){\psfig{figure=ig4.ps,height=4.8cm,angle=-90,clip=}}
\end{picture}
\caption[]{{\bf a:} Comparison of the X-ray luminosity log(L$_{\rm X}$) 
with the \lxb-ratio for colliding starburst galaxies. 
We include NGC\ 2623 and AM\ 1146-270 (Read \& Ponman 1998), 
Arp\ 284 (Papaderos \& Fricke 1998) and NGC\ 7252 (Fritze v.-Alvensleben et al. 1998). 
The samples of David et al. (1992) and Read et al. (1997) are shown by
dots and open circles, respectively.
{\bf b:} \lirb\ vs. \lxb-diagram. 
{\bf c:} \lirb- vs. e/p-ratio.
{\bf d:} Evolution of the \lxb\ and \lirb\ ratios along a tentative merging
sequence.}
\label{IG4}
\end{figure}
We arranged our 22 objects tentatively along a qualitative merger sequence
mainly being guided by the interaction strength and morphology of the systems
(distances, disturbances, tails, bridges, brightness profiles). 
In Fig.\ 4d we plot \lxb\ and \lirb\ along this sequence.

Three phases can be roughly distinguished.
On the l.h.s are located the early, detached systems like Arp\ 284 
which experience a first interaction-induced starburst while the 
r.h.s is populated by the advanced mergers with strong nuclear 
starbursts, like NGC\ 6240.
The intermediate portion represents those systems which are just starting
the merging process like the Antennae (NGC\ 4038/39).
The strong decline in the FIR- and X-ray luminosity 
on the right end represents the post merger stage and 
the transition of the merger into an elliptical galaxy as 
seen in NGC\ 7252.\\

Observations of systems populating the intermediate phase 
show an increasing pile-up of obscuring material (mainly molecular gas) 
enshrouding the starburst region. 
In this phase the FIR luminosity starts to rise until 
dominating the bolometric luminosity for objects beyond 
NGC\ 6240 on the sequence.
\section{Discussion and conclusions}
The evolution of individual systems along this sequence may 
strongly depend on the intrinsic properties of the colliding galaxies
as has been shown in dissipative numerical simulations by 
Mihos \& Hernquist (1996; cf. their Fig.\ 10): 
The disk/halo systems suffer their main burst early in the encounter and
hardly burst again, while the disk/bulge/halo systems manage to retain 
their gas almost until complete merging, in order then to suffer a strong, 
short-lived burst much later than the bulgeless systems. 
For the merged systems it is difficult to verify a posteriori 
whether the original interaction partners were bulge systems or not.
These important parameters are not directly known.
Dissipative numerical simulations with very high particle numbers 
may help here. 
Furthermore, it is not well understood, how the formation 
of non-thermal activity (AGNs) fits into the merger sequence. 
The sample investigated here 
shows the occurrence of AGNs in all major phases of the sequence.
Also here very high-resolution gas/star numerical simulations
are required to investigate the inflow conditions at the centers of 
interacting/merging systems.\\
The extended X-ray contribution e/p does not show 
any conspicuous evolution along the merger sequence.
This extended component is in all phases related to the
starburst activity and the shock-heating of the ambient gas by 
starburst winds.
This mechanism is certainly not responsible for the extended 
X-ray emission observed in the halos of elliptical galaxies (Fabbiano 1989).
In fact, in the young merger remnant NGC\ 7252 an extended X-ray 
halo is not yet present.

Around such systems, however, ample mechanical energy is stored 
in the gravitationally bound tidal material.
During the ensuing passive evolution this material will fall back 
onto the main body of the remnant and will thereby form an extended X-ray halo as 
seen in Ellipticals within a couple of Gyrs.
Model calculations for this process will be important (cf. Hibbard \& Mihos 1995).\\
The example of Arp\ 284 (Papaderos \& Fricke 1998) 
may indicate that off-nuclear X-ray sources may form 
quite early during interactions in the shocked \hi-gas.
Such sources will later-on be smeared out  
and may contribute to halo formation at a stage prior 
to completed merging.
More examples of this type should be found.\\ 
Understanding of the merger sequence, e.g. the evolution of 
global parameters like the luminosities in different energy bands 
and the morphology of extended emission in successive stages of merging
is still in a rudimentary state.  
The complexity, in particular regarding the generation of
X-rays, is enormous as seen from the detailed results for individual 
objects (cf. Papaderos \& Fricke 1998, Fritze, Read \& Ponman 1998).
However, first patterns seem to emerge.

\vspace*{-2mm}
\begin{acknowledgements}
We thank W.\ Pietsch and K.\ Bischoff for observing an U-band frame of NGC\ 6240.
P.\ Papaderos thanks K.\ Bischoff for assistance during observations
at Calar Alto.
This work was supported by the Deutsche Agentur f\"{u}r Raumfahrtangelegenheiten 
(DARA) grant 50\ OR\ 9407\ 6.
\end{acknowledgements}

\end{document}